\newcommand{\footurl}[1]{\footnote{\url{#1}}}

\documentclass[sigconf,nonacm=true]{acmart}

\author{Juha Nurmi}
\orcid{0000-0003-3071-9027}
\affiliation{
\institution{Tampere University}
\city{Tampere}
\country{Finland}
}
\email{juha.nurmi@tuni.fi}
\email{juha@cyberintelligencehouse.com}

\author{Mikko Niemel\"{a}}
\orcid{0000-0001-9321-5235}
\affiliation{
\institution{Cyber Intelligence House}
\city{Singapore}
\country{Singapore}
}
\email{mikko@cyberintelligencehouse.com}

\author{Billy Bob Brumley}
\orcid{0000-0001-9160-0463}
\affiliation{
\institution{Tampere University}
\city{Tampere}
\country{Finland}
}
\email{billy.brumley@tuni.fi}

\newcounter{rqcounter}
\makeatletter
\newcommand{\rqcounterautorefname}{RQ\@gobble}

\newcommand{\INFECTIONS}{Mal\-ware\-In\-fec\-tion\-Set}
\newcommand{\VICTIMS}{Vic\-tim\-Ac\-cess\-Set}
\newcommand{\COMPROMISE}{Ac\-countAc\-cess\-Set}

\newcommand{\KEYWORDS}{%
Anti-virus Software;
Computer Hacking;
Malicious Program Traits;
Intrusion Detection}

\title[Malware Finances and Operations]{Malware Finances and Operations: a Data-Driven Study of the Value Chain for Infections and Compromised Access}

\begin{abstract}
We investigate the criminal market dynamics of infostealer malware
and publish three evidence datasets on malware infections and trade.
We justify the value chain between illicit enterprises using the datasets,
compare the prices and added value,
and use the value chain to identify the most effective countermeasures.

We begin by examining infostealer malware victim logs shared by actors on hacking forums,
and extract victim information and mask sensitive data to protect privacy.
We find access to these same victims for sale at Genesis Market.
This technically sophisticated marketplace provides its own browser to access victim's online accounts.
We collect a second dataset and discover that 91\% of prices fall between 1--20 US dollars, with a median of 5 US dollars.

Database Market sells access to compromised online accounts.
We produce yet another dataset,
finding 91\% of prices fall between 1--30 US dollars, with a median of 7 US dollars.
 \end{abstract}
\copyrightyear{2023}
\acmYear{2023}
\setcopyright{rightsretained}
\acmConference[ARES 2023]{The 18th International Conference on Availability, Reliability and Security}{August 29-September 1, 2023}{Benevento, Italy}
\acmBooktitle{The 18th International Conference on Availability, Reliability and Security (ARES 2023), August 29-September 1, 2023, Benevento, Italy}\acmDOI{10.1145/3600160.3605047}
\acmISBN{979-8-4007-0772-8/23/08}

\begin{document}

\maketitle

\section{Introduction}\label{sec:intro}

Malicious software, or malware, is a persistent executable on a victim system controlled by an attacker \cite{DBLP:journals/cj/GregioAFGJ15}.
Infostealer malware steals personal information from the victim \cite{DBLP:conf/acsac/KaczmarczyckGIP20}.
This typically includes (but is not limited to) system data, login credentials, browser cookies,
payment card information, and cryptocurrency wallets \cite{campobasso2020impersonation}.
Existing literature classifies malware in a tree-like structure, and there are distinct
families of infostealer malware \cite{DBLP:conf/sigsoft/ChristodorescuJK07,DBLP:journals/compsec/ParkRS13,esparza2019understanding,vandepth}.

The five most popular infostealers are Taurus, Raccoon, Azorult, Vidar, and RedLine \cite{Cybersixgill}.
We also found active Kpot and Arkei malware family victims, and all of them are available for sale \cite{vmware}.
After purchasing malware control software, the adversary convinces victims to download malicious code and become infected.
Infostealers communicate with a Command \& Control (C\&C) Server,
which remotely controls the victim's computer \cite{vmware,vandepth}.

Protection software cannot detect all intrusions by malware \cite{DBLP:conf/ccs/Axelsson99,DBLP:conf/uss/KolbitschCKKZW09},
because code obfuscation methods hide the infection \cite{DBLP:journals/virology/BorelloM08}.
Predator the Thief, for example, is an infostealer available on hacking forums,
and is constantly updated to include a variety of mechanisms to evade detection \cite{Predator2020}.
Updated stealth mechanisms make detection difficult for anti-virus software \cite{DBLP:conf/sp/ChristodorescuJSSB05,DBLP:series/ais/FredriksonCGJ10}.

Unethical software developers provide infostealer control software via Malware-as-a-Service (MaaS) \cite{raccoon2022},
which is a form of Cybercrime-as-a-Service (CaaS).
In 2022, prices range from 75 US dollars per week,
to 125--200 US dollars per month,
to 1,000 US dollars for an unlimited lifetime subscription for improved versions of RedLine \cite{raccoon2022,meta2022}.
Subscribing threat actors gain access to an administration panel enabling them to customise the malware, retrieve stolen data,
and create new malware builds \cite{raccoon2022}.

Malicious actors infect victims with infostealer malware using (most frequently)
phishing emails, cracked and pirated software, game cheating packages, browser extensions,
and cryptocurrency-related software \cite{campobasso2020impersonation,sansmeta2022}.
In 2022, for example, one infostealer spread via spam emails that included an Excel spreadsheet containing macros
that downloaded a persistent executable \cite{sansmeta2022}.

The malware network owner profits from victim data
by gaining access to the compromised services or by selling the victim data to other criminals who profit from it \cite{vandepth,esparza2019understanding}.
A study from 2016 indicates,
``Cybercriminals steal access credentials to webmail accounts and then misuse them for their own profit,
release them publicly, or sell them on the underground market'' \cite{DBLP:conf/imc/OnaolapoMS16}.
For example, between January 2021 and March 2022, one particular marketplace sold 4,368,909 units of victim data \cite{Cybersixgill}.

\subsubsection*{Research questions.} We propose the following research questions related to the previously described actors in the malware ecosystem.

\begin{description}%
    \item[\autoref*{rq:valuechain}.]\refstepcounter{rqcounter}\label{rq:valuechain}%
    How do criminal enterprises sell malware, victims, and online account access?
    \item[\autoref*{rq:device}.]\refstepcounter{rqcounter}\label{rq:device}%
    How are the prices determined for compromised devices?
    \item[\autoref*{rq:accounts}.]\refstepcounter{rqcounter}\label{rq:accounts}%
    Compared to access to a victim's device, what is the cost of online account access?
\end{description}
\subsubsection*{Contributions.}
These actors form illicit markets, and we study them (in part) using economics nomenclature:
``variables that can be observed in, or derived directly from, markets (for example, prices of publicly traded securities and interest rates)''
(A926 IFRS 17 Insurance Contracts manual\footurl{https://www.ifrs.org/content/dam/ifrs/publications/pdf-standards/english/2021/issued/part-a/ifrs-17-insurance-contracts.pdf}).
Our interests resemble those of economics literature:
``The interest in market factors is threefold:
(1) to identify the factors that influence a product's or service's demand,
(2) determine the relationship between the factor and the product or service,
and (3) forecast that market factor for future years.'' \cite[p.\ 52]{winston2013market}.

\textbf{Contribution 1: Release data describing the value chain.} For our research, we collect, curate, and model three datasets, which we publish under the CC BY 4.0 licence.
We represent each victim using a single JavaScript Object Notation (JSON) data file.
Data sources provide sets of victim JSON data files from which we extract the essential information
and omit Personally Identifiable Information (PII).

\subsubsection*{\INFECTIONS{}: Underground hacking forums for malware logs.}
RaidForums%
\footurl{https://raidforums.com/}
operated as a forum for unethical hackers from 2015 until 2022, when the FBI seized the domain.
The website was distributing data breaches and malware tools.
We use RaidForums as a starting point to track criminal profiles,
and then add Telegram and Discord channels based on their posts.
This enables us to follow the content criminals share in a broader fashion.

We discover (and, to the best of our knowledge, document scientifically for the first time)
that malware networks appear to dump their data collections online.
We speculate this could possibly be due to one or more of the following factors.
(i) They exploited the victims and hid their tracks by attracting others to exploit the victims on a large scale as well.
For example, for at least a decade on pastebin\-.com, cybercriminals have released publicly accessible lists to compromised targets \cite{bbc2012}.
(ii) By offering free example data, they showcase their skills and specialisation.
For example, spammer botnets distribute free samples and display their capabilities in this manner \cite{DBLP:conf/leet/Stone-GrossHSV11,DBLP:conf/imc/OnaolapoMS16}.
(iii) A rival adversary leaked information to harm the operations of another cybercriminal;
see \cite{intsights} for an example case from 2019.

In this study, we utilise 245 malware log dumps from 2019 and 2020 originating from 14 malware networks.
\INFECTIONS{} contains 1.8 million victim files, with a dataset size of 15 GB\null.

\textbf{Contribution 2: Sale of compromised devices.} To answer \autoref{rq:valuechain}, we demonstrate how Infostealer malware networks sell access to infected victims.

\subsubsection*{\VICTIMS{}: Genesis Market for compromised access.}
Genesis Market%
\footurl{https://genesis7zoveavupiiwnrycmaq6uro3kn5h2be3el7wdnbjti2ln2wid.onion/}
is a cybercriminal marketplace specialising in the sale of web browser data from malware victims
(see ``Buying Bad Bots Wholesale: The Genesis Market'' report in 2021\footurl{https://www.netacea.com/wp-content/uploads/2021/04/Genesis_market_report_2021.pdf}).
The marketplace operates within the anonymous Tor network but also through the clear web;
the \texttt{genesis\-.market} domain was accessible until April 2023, before the FBI seized it.

Genesis Market focuses on user-friendliness and continuous supply of compromised data.
Marketplace listings include everything necessary to gain access to the victim's online accounts,
including passwords and usernames,
and also detailed information which provides a clone of the victim's browser session.
Indeed, Genesis Market simplifies the import of compromised victim authentication data into a web browser session.

To utilise the stolen data in a user-friendly way, Genesis Market offers both a web browser plugin and a dedicated web browser:
the Genesis Security Plugin, a standard browser plugin, or the Genesium Browser,
a Chromium-based browser built specifically for Genesis Market users, featuring the Genesis Security Plugin pre-installed.
Once a customer purchases a victim's complete access information,
browser technology enables the customer to access the victim's stolen online profiles and authenticated sessions.

\textbf{Contribution 3: Victim prices.} To answer \autoref{rq:device}, we measure the prices on Genesis Market and how compromised device prices are determined.

To build our dataset \VICTIMS{},
we crawled the website between April 2019 and May 2022.
We use the collected web pages offering the resources for sale
to investigate market factors -- in particular those that influence the sale price.
\VICTIMS{} contains 0.5 million victim files,
with a dataset size of 3.5 GB\null.

\subsubsection*{\COMPROMISE{}: Database Market.}
Database Market%
\footurl{http://database6e2t4yvdsrbw3qq6votzyfzspaso7sjga2tchx6tov23nsid.onion/}
specialises in selling credentials to compromised online accounts.
The marketplace operates inside the anonymous Tor network.
Vendors offer their goods for sale, and customers can purchase them with Bitcoins.
The marketplace sells online accounts,
such as PayPal and Spotify, as well as private datasets,
such as driver's licence photographs and tax forms.

To build our dataset \COMPROMISE{},
we crawled the website between November 2021 and June 2022.
\COMPROMISE{} contains 33,896 victim files,
with a dataset size of 400 MB\null.

\textbf{Contribution 4: Account prices.}
To answer \autoref{rq:accounts}, we measure the prices on Database Market and illustrate an increase in value:
an attacker with access to a single victim can sell access to individual accounts at a higher price.

\textbf{Contribution 5: Value chain.}
The contributions reveal a broader perspective:
we draw the value chain for the malware economy from our findings and related prices in the literature.

\subsubsection*{Structure.}
\autoref{sec:background} discusses previous malware-related work in the context of CaaS, including economic aspects.
\autoref{sec:methods} presents our methods, and
\autoref{sec:instructions} shares the three datasets and provides a step-by-step tutorial to allow anyone to reproduce our findings.
Consequently, \autoref{sec:results} provides our primary contributions:
we investigate the prices
and the value chain between illicit enterprises.
\autoref{sec:quality} establishes coverage metrics to measure the quality and extensiveness of our datasets.
\autoref{sec:privacy} discusses ethical and privacy considerations we account for in our work.
We conclude our research in \autoref{sec:conclusion}.
\section{Related work}\label{sec:background}

\subsubsection*{Cybercrime economics.}
In prior research, the cybercrime value chain has been established in \textit{Framing Dependencies Introduced by Underground Commoditization} \cite{DBLP:conf/weis/ThomasHWBGHKMSV15},
however this value chain is justified with circumstantial evidence and the data are not interconnected,
as the research points out:
``the future of black market research lays in developing a better understanding of the tenuous connections between the actors in this space \dots
we note that research into many of these profit centers is sorely lacking: we rely on industry and government estimates of profit in the absence of methodologically sound measurements
\dots research contributions must shift from analysis based off anecdotal evidence to thorough investigations of the ecosystem''.
In our study, we address this gap in the literature and
we establish robust connections between the malware networks and the illicit markets.

Already in 2011, \citet{DBLP:conf/leet/Stone-GrossHSV11} identify financial variables and the specialisation of cybercriminals.
The authors describe how botnet operators sell their services, while other vendors only sell email address lists:
``spam-as-a-service can be purchased for approximately \$100--\$500 per million emails sent''
and
``Rates for one million email addresses range from \$25 to \$50''.

\citet{tmp:Broadhurst18} study MaaS offerings on the illicit Dream Market
from 2017 to 2018.
During this period, they calculate that 47 AUD is the average price for
a single set of compromised credentials.

\citet{DBLP:journals/csur/HuangSM18} propose a conceptual model describing CaaS.
The authors identify an impressive 24 families of CaaS,
including pricing models and estimated costs.
Our dataset \INFECTIONS{} falls under their categorisation
Personal Profile as a Service (PPaaS),
for which they provide a representative price of 4--20 US dollars per record
in 2018 \cite[Tbl.\ 1]{DBLP:journals/csur/HuangSM18}.

\citet{tmp:Meland19} study 11 dark web illicit markets offering CaaS.
They explicitly omit compromised credentials from their study,
since their focus is on the sale of offensive services \cite[Sect.\ 3]{tmp:Meland19}.

\citet{DBLP:conf/ecrime/BhaleraoASAM19} study CaaS using machine
learning techniques to perform supply chain detection.
They validate their techniques using the same underground forum
as \citet{DBLP:conf/imc/VuHPCCSH20}.

Lastly, \citet{uss/Campobasso2023} crawl the Russian cybercrime market
that provides user impersonation profiles and estimate a daily trade volume of up to 700 profiles,
equating to daily sales of up to 4,000 US dollars.

\subsubsection*{Security dataset classification.}
\citet{DBLP:conf/uss/ZhengRCTM18} study security-related datasets
to provide a taxonomy, as well as to understand the scholarly
ramifications of releasing said datasets. In their classification,
our datasets falls under the ``Attacker-Related'' category, and
furthermore the ``cybercrime activities'' subcategory, since
they contain ``information on the infrastructure and operations
used by malicious actors to perpetrate attacks'' \cite[Sect.\ 3]{DBLP:conf/uss/ZhengRCTM18}.
While the authors give other examples from this class including data
collected from illicit marketplaces inside the Tor
network \cite{DBLP:conf/www/Christin13,tmp:2015:jdrug:dolliver,tmp:2016:victims:dolliver,tmp:2017:dep:nurmi},
ours is the first study dedicated to the malware ecosystem.

\subsubsection*{Gathering cybercrime data.}
\citet{DBLP:conf/imc/MotoyamaMLSV11} study data from six underground forums
dedicated to the sale of illicit goods and services.
They focus on social network aspects, including connectivity, reputation,
and community policies.
Furthermore, they study these marketplaces to determine how the different
social network aspects affect market performance.
One of these forums (Carders, German speaking) specialised in stolen
credit card numbers,
yet interestingly roughly 3\% (334/9923) of the sale listings were victim
logs \cite[Tbl.\ 5]{DBLP:conf/imc/MotoyamaMLSV11}.
The original data source for their study was leaked SQL
databases \cite[Sect.\ 3]{DBLP:conf/imc/MotoyamaMLSV11},
sharing conceptual similarities with the original data source for
our \INFECTIONS{}.

\citet{DBLP:conf/cycon/0002FSELL19} build a tool called BlackWidow to
monitor illicit forums.
The tool is also capable of machine learning-based information extraction.
The authors apply their tool to seven different forums,
subsequently analysing the cross-forum social network aspects.

\citet{DBLP:conf/www/PastranaTHC18} propose CrimeBot, a framework for
scalable data collection from illicit online forums. The authors
apply CrimeBot to produce their CrimeBB
dataset \cite[Sect.\ 4]{DBLP:conf/www/PastranaTHC18},
containing 48 million posts and their associated metadata.
CrimeBB continues to grow,
now approaching 100 million posts in 2022 \cite{DBLP:conf/eurosp/PeteHCVGHAB22}.

\subsubsection*{Cybercrime data creates new research.}
As CrimeBB grows, it increasingly fosters academic studies for those
with access to the data through a legal agreement with the University of
Cambridge.\footnote{\url{https://www.cambridgecybercrime.uk/process.html}}
The following CrimeBB usage examples are relevant to our research
and motivate us to publish datasets for the benefit of other researchers.

\citet{DBLP:conf/raid/PastranaHCB18} utilise CrimeBB data to characterise the key individuals that are
likely to engage in major cybercriminal activities, aiding law enforcement.
In a similar vein with a focus on individuals,
\citet{DBLP:conf/eurosp/PeteHCB20} analyse social network aspects
of six dark web forums and use graph-theoretic similarities to identify properties
in individual users that lead to these actors taking different forum roles.
In a CrimeBB followup study,
\citet{tmp:Collier20} investigate cybercrime infrastructure work
and show that botnet administration tasks are a burden
and a weak point in the cybercrime ecosystem,
suitable for law enforcement targeting.
\citet{DBLP:conf/imc/VuHPCCSH20} analyse the
reputation properties in one particular illicit marketplace,
and \citet{DBLP:conf/ecrime/Bermudez-Villalva21}
compare dark web vs.\ deep web markets.

Lastly, and most relevant to our study, \citet{tmp:Akyazi21}
use the CrimeBB data to analyse various CaaS
offerings on one particular illicit marketplace.
The vast majority of these offerings were related to botnets, reputation
boosting, and traffic flooding.

\section{Methods}\label{sec:methods}

\begin{figure*}[ht!]
\centering
\includegraphics[width=\textwidth]{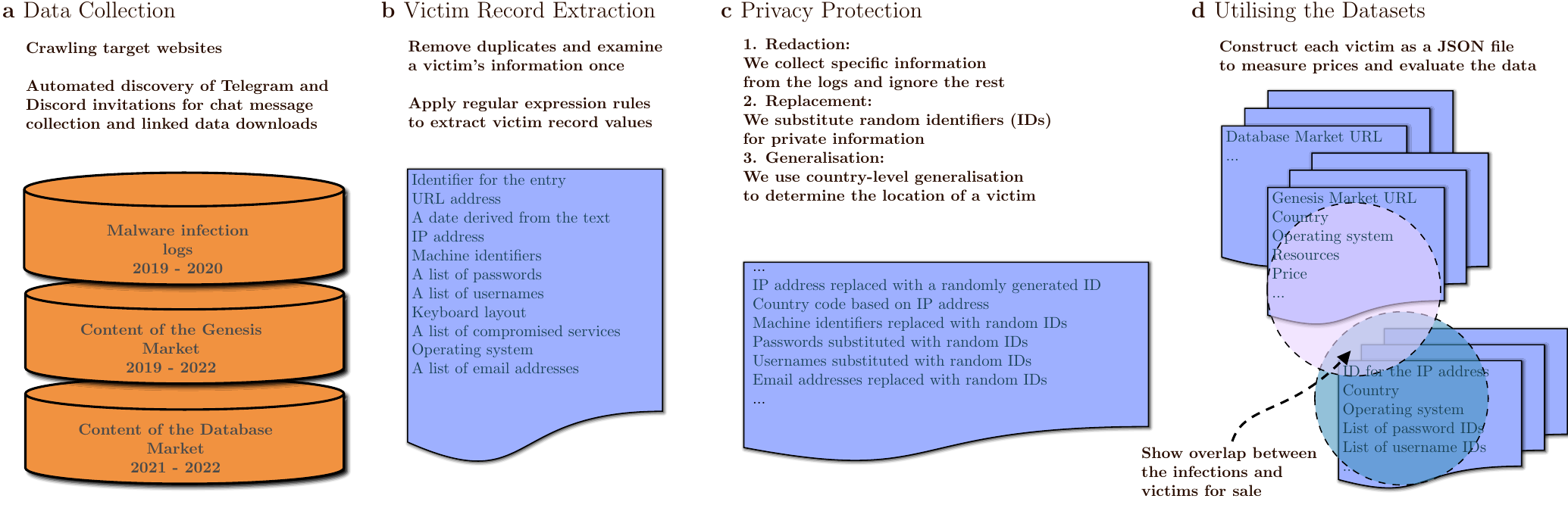}
\Description[Malware ecosystem monitoring]{A summary of our methods for automated malware ecosystem monitoring.}
\caption{%
\underline{(a)} Our crawlers authenticate on RaidForums, Genesis Market, and Database Market for data collection.
Our bots follow messages by joining the Discord and Telegram channels mentioned on RaidForums.
They download linked resources, including malware infection log files.
\underline{(b)} To extract victim records from the data sources, we use regular expressions.
\underline{(c)} We apply redaction, replacement, and generalisation to omit PII\@.
\underline{(d)} Finally, we create JSON victim record sets for pricing measurements, demonstrate overlap between infections and for-sale victims, and evaluate our data.%
}\label{fig:methods}
\end{figure*}

\autoref{fig:methods} provides a summary of our automated malware ecosystem monitoring:
web crawling, chat message collection, and linked resource downloads.
We process the raw data into victim record files while protecting the user's privacy.
The privacy protection measures replace the same input with the same ID\@.
For instance, throughout the dataset, the same IP address has the same ID\@. Similarly,
if there are passwords for services but the passwords are identical, the replacement IDs are identical.
We curated together datasets to measure malware infections, the sale of access to victims,
and the reselling of online accounts (see \autoref{sec:instructions} on how to download the datasets and scripts and replicate the results).

\subsection{Data extraction for \INFECTIONS{}}

Our web crawler authenticates with RaidForums in order to gather pages and extract Discord and Telegram invitations.
Then, our bots join channels on Discord and Telegram to monitor messages.
The bots visit the shared links and download resources,
including log files containing malware
(if this is not feasible, the resources are reported for manual human involvement).

\begin{figure}
\centering
\includegraphics[width=\linewidth]{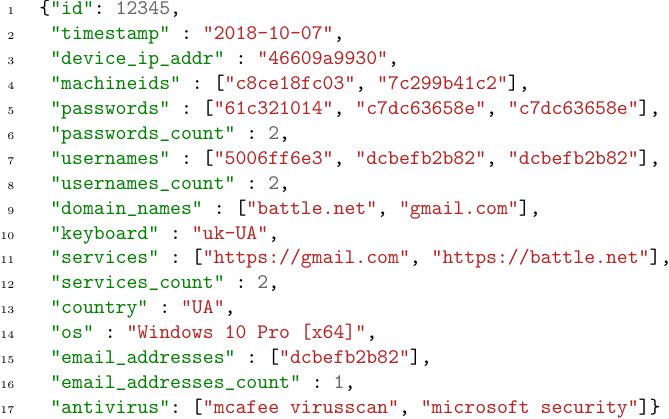}
\Description[Infostealer victim entry]{An example JSON file represents a victim entry.}
\caption{An example JSON file from \INFECTIONS{} depicting victim data.}\label{fig:j_victim}
\end{figure}

We utilise 245 malware log dumps from 2019 and 2020 originating from
AZORult,
KPOT,
Raccoon,
Qulab,
Masad,
Arkei,
Oski,
Taurus,
Redline,
Mystery,
Ficker,
Predator,
Ducky,
and Vidar
malware networks.
The total size of these original malware log files is 30 GB\null.
\autoref{fig:j_victim} is an example of a victim record file.

The malware logs include system information, login credentials to online services, user information, credit card numbers, screenshots of victim desktops,
operating system, keyboard languages, IP address, geolocation, hardware specifications, process list of active software,
browser information and cookies, cryptocurrency wallets, victim files, automatically filled site forms, gaming software configurations, chat service configurations,
clipboard data, installed software, browsing and download history, and mail, VPN, RDP, and FTP account information.

We structured and cleaned the data, aggregating the following fields from the logs
to build our dataset \INFECTIONS{}:
(1) Identifier for the entry;
(2) A date derived from the logs;
(3) IP address replaced with a randomly generated ID\@;
(4) Machine identifiers replaced with random IDs;
(5) A list of passwords we substituted with random IDs;
(6) Total number of one-of-a-kind passwords;
(7) A list of usernames we substituted with random IDs;
(8) Total number of distinct usernames;
(9) All domain names in the log, including browser history;
(10) Keyboard layout;
(11) A list of compromised services;
(12) Count of compromised services;
(13) Country code which is primarily based on IP address;
(14) Operating system;
(15) A list of email addresses replaced with random IDs;
(16) Total number of unique email addresses;
(17) Possible endpoint protections.

We remove email addresses, IP addresses, and payment card numbers from raw text fields, such as those used to list compromised services.
This means that service count (number of distinct services) may exceed the total number of services listed.
In that case, we remove services which contain private information from the list.

\subsection{Data extraction for \VICTIMS{}}

\begin{figure}
\centering
\includegraphics[width=\linewidth]{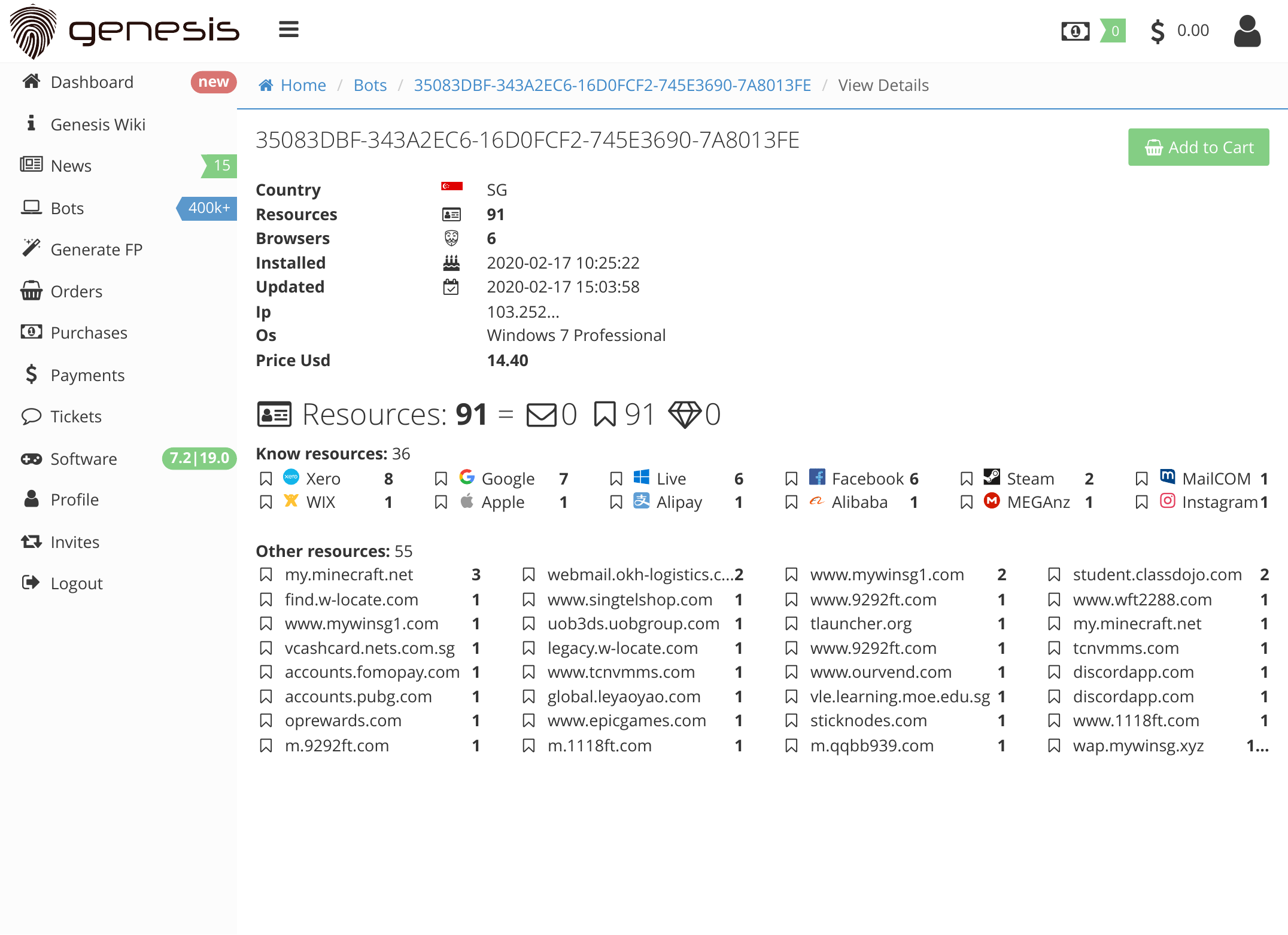}
\Description[Genesis Market]{The infostealer malware adds compromised data to the Genesis Market for sale.}
\caption{The infostealer malware adds compromised data to the Genesis Market for sale.}\label{fig:genesis}
\end{figure}

We crawled the Genesis Market between April 2019 and May 2022, collecting the web pages offering the resources for sale.
\autoref{fig:genesis} is an example of a victim information page.

\begin{figure}
\centering
\includegraphics[width=\linewidth]{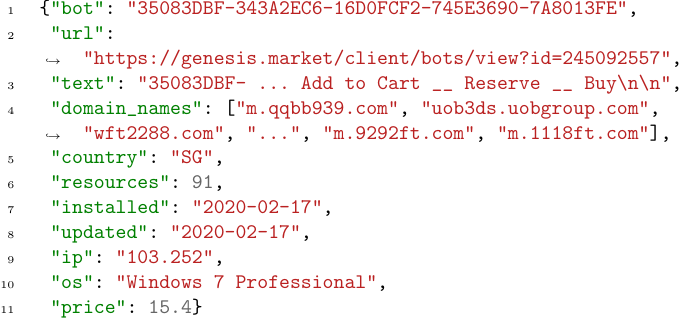}
\Description[Genesis Market victim entry]{An example JSON file of compromised data on the Genesis Market.}
\caption{An example JSON file from \VICTIMS{} which represents one victim.}\label{fig:j_victim_genesis}
\end{figure}

Since the victim page is continuously updated,
we use the oldest page as the reference entry (or entry of record) in \VICTIMS{}.
As such, \autoref{fig:j_victim_genesis} is an example of a file.
This record corresponds to the page in \autoref{fig:genesis},
however, due to different capture dates, the price has increased from 14.40 to 15.50 US dollars.

We consolidated the following fields to build our dataset \VICTIMS{}:
(1) Bot identifier from the webpage;
(2) URL address;
(3) Text of the webpage;
(4) All domain names on the webpage;
(5) Country code;
(6) Total number of resources;
(7) The malware installation date;
(8) Date of victim information update;
(9) Partial IP address;
(10) Operating system; and
(11) Price of the access.

\subsection{Data extraction for \COMPROMISE{}}

\begin{figure}
\centering
\includegraphics[width=\linewidth]{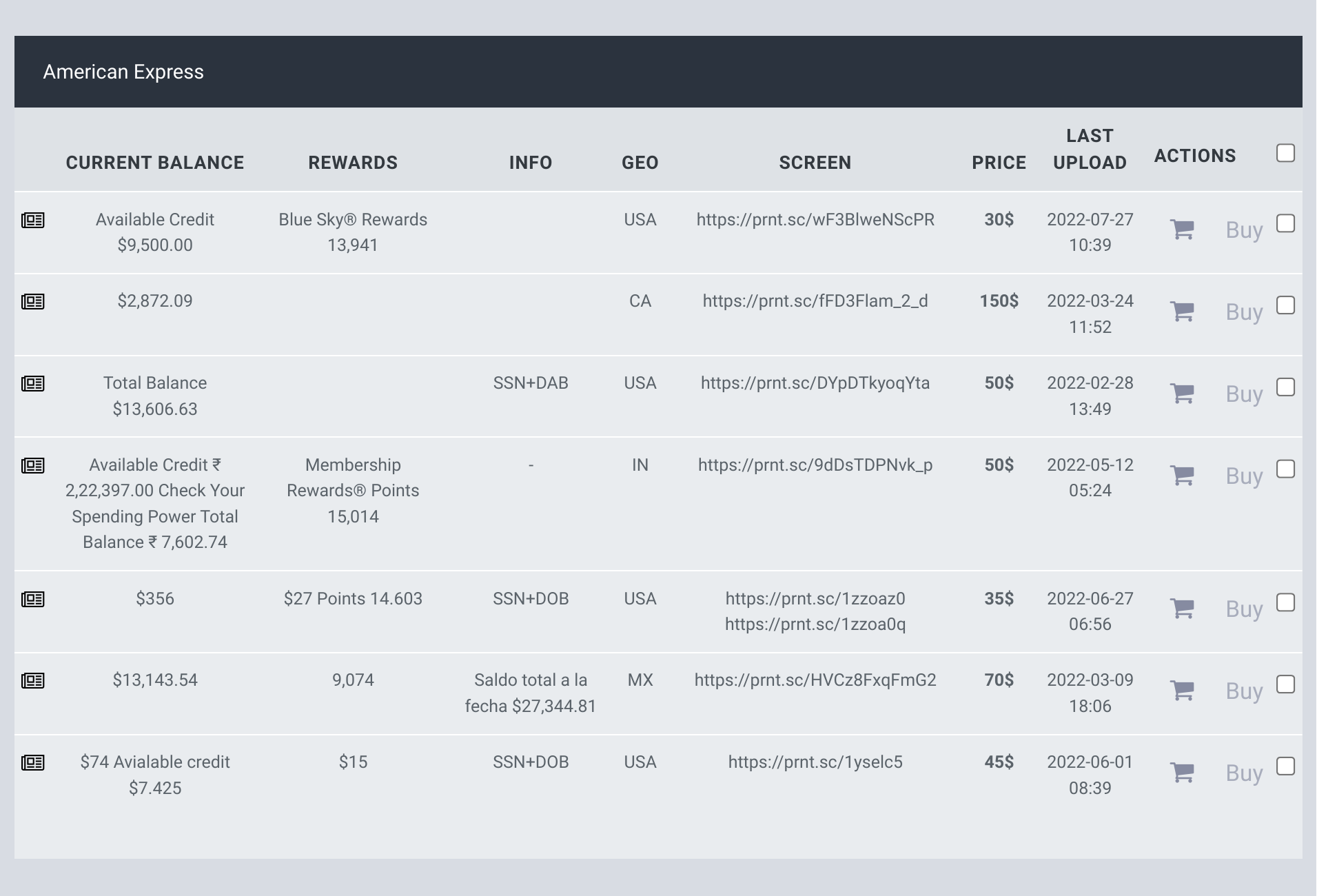}
\Description[Database market]{Compromised data for sale on Database market.}
\caption{Compromised data for sale on Database market.}\label{fig:database}
\end{figure}

We crawled the website between November 2021 and June 2022, collecting the web pages offering the credentials for sale.
\autoref{fig:database} is an example of a vendor advertising compromised credentials for sale on Database Market.

\begin{figure}
\centering
\includegraphics[width=\linewidth]{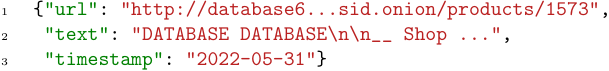}
\Description[Database market victim data]{An example JSON file of compromised data on the Database market.}
\caption{An example JSON file from \COMPROMISE{} which represents one page.}\label{fig:j_victim_database}
\end{figure}

\autoref{fig:j_victim_database} is an example of a file,
where we populated the following fields:
(1) URL address of the webpage;
(2) Text of the webpage; and
(3) Data collection date.

Because vendors can style advertisements in any way they like,
it is not possible to accurately collect price and sale item information across the full \COMPROMISE{} in an automated way.
We manually inspected a random sample of 599 out of 33,896 pages on Database Market.
We carefully verified that our chosen subset contains individual accounts with price information and no other products for sale, such as merged accounts, company databases, or stolen software.
Then, we created an extraction script for these 599 pages to evaluate the prices.

\subsection{Relationship between malware infections and victims for sale on Genesis Market}

To illustrate the value chain, we compare the malware infections with the records for sale on Genesis Market by cross-referencing \INFECTIONS{} with \VICTIMS{}.
We are identifying unique users based on service and country data, then comparing their presence in both datasets.

For instance,
the victim information for sale on Genesis Market,
present in \VICTIMS{} (ID=245092557, 245092557.json)
and depicted in \autoref{fig:j_victim_genesis},
has details matching a record from \INFECTIONS{} (ID=487238, 0487238.json).
This Singaporean victim has the following rare compromised services:
\texttt{m\-.qqbb939\-.com},
\texttt{either\-.io},
\texttt{wft2288\-.com},
\texttt{m\-.9292ft\-.com},
\texttt{m\-.1118ft\-.com},
\texttt{uob3ds\-.uobgroup\-.com},
\texttt{legacy\-.w-locate\-.com}, and
\texttt{tcnv\-mms\-.com}.

The matching is as follows:
(i) select the victims that have at least five completely unique services based on the domain names;
(ii) calculate the intersection of three or more of these unique domain names between the datasets;
and (iii) check that the countries are identical.

The matching criteria identify users who are unique based on the services they use
(they use uncommon services that other users do not)
and who are present in both datasets.
As a result,
the tight matching criterion finds only the most unique victims in both datasets,
not all of the victims in both datasets.
\section{Datasets to replicate our results}\label{sec:instructions}

Using our well-documented scripts, it is straightforward to reproduce our findings.
You can download our datasets, Python scripts, and tutorials on the Zenodo service for Open Science\footurl{https://doi.org/10.5281/zenodo.8047204}.
It takes an estimated 1 hour of human time and 3 hours of computing time to duplicate our key findings from \INFECTIONS{} (malwarevictims.tar.gz, 1.7 GB);
around one hour with \VICTIMS{} (genesisvictims.tar.gz, 335 MB);
and minutes to replicate the price calculations using \COMPROMISE{} (database.tar.gz, 7 MB).
\section{Results}\label{sec:results}

\textit{\autoref*{rq:valuechain}: How do criminal enterprises sell malware, victims, and online account access?}

We discovered information-stealing malware logs from victims of the Taurus, Raccoon, Azorult, Vidar, RedLine, Kpot, and Arkei families.
These MaaS controlling software are for sale on underground markets with prices ranging from 75 to 200 US dollars per week and 125 to 200 US dollars per month.
After acquiring an MaaS subscription, the adversary convinces victims to download malicious code, causing them to become infected.

\subsection{Malware targets Windows}

The infostealers target users of Microsoft's Windows operating system. Windows 10, Windows 7,
and Windows 8 are the main infected operating systems according to \INFECTIONS{} and \VICTIMS{}.
Victims in \VICTIMS{} employ one of the following Windows operating systems:
76\% (365,676/480,994) use Windows 10, 7\% (32,989/480,994) use Windows 8, and 16\% (77,170/480,994) use Windows 7.
We extracted \INFECTIONS{} from heterogeneous text data logs in which the operating system information is sometimes absent, incorrect, or described in multiple ways;
nonetheless, if we search os:NT* or os:Windows*, we obtain 62\% (1,115,263/1,809,384)
of the infection victims.

Stealth mechanisms are making malware difficult to detect for anti-virus software.
According to \INFECTIONS{}, some users have installed popular
anti-virus software:
Avast, Mcafee, Symantec, Microsoft Security Essentials, Bitdefender, Microsoft/Windows Defender, F-secure anti-virus, Zonealarm anti-virus, and Kaspersky anti-virus.
The infection occurred, regardless.

\subsection{Credentials and password reuse}

Password reuse is a common problem for security:
We study credential reuse with \INFECTIONS{}, and find that 50\% of victims have 1--8 unique passwords.

\begin{figure}[tb]
\centering
\includegraphics[width=\linewidth]{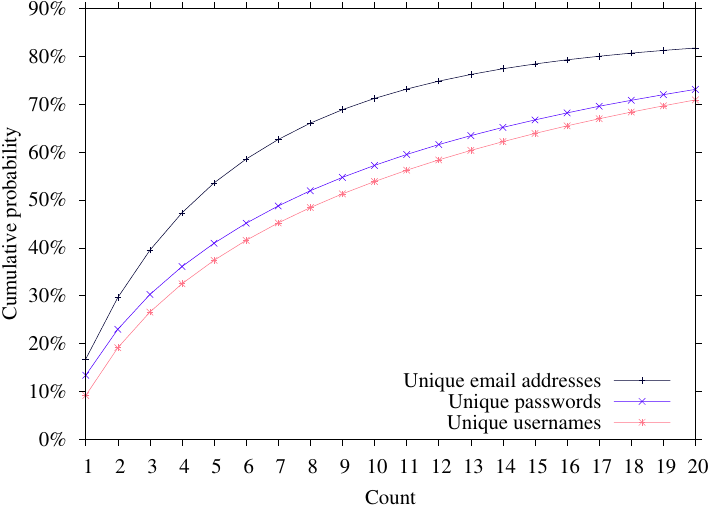}
\Description[Credentials]{Cumulation of unique passwords, usernames, and email addresses.}
\caption{Cumulation of unique passwords, usernames, and email addresses for each victim in \INFECTIONS{}.}\label{fig:passwords_usernames_emails}
\end{figure}

Furthermore, \autoref{fig:passwords_usernames_emails} also shows that the majority of victims have few distinct usernames.
In total, 9\% of victims have a single unique username across services, while 10\% have two.
The majority of victims have between 1--9 usernames and between 1--5 unique email addresses.

If we perform linear regression on the number of unique usernames and unique passwords across all victims,
we discover that a typical user has 1.12 times as many unique usernames as unique passwords:
\begin{displaymath} Usernames = 1.12*Passwords + 1.78 \end{displaymath}

Investigating the correlation between the number of unique credentials and compromised services similarly with linear regression, we obtain:
\begin{displaymath} EmailAddresses = 0.11*Services + 3.67 \end{displaymath}
\begin{displaymath} Usernames = 0.36*Services + 6.19 \end{displaymath}
\begin{displaymath} Passwords = 0.31*Services + 4.43 \end{displaymath}

This means that a typical user creates a new unique password and username for every third service.

\subsection{Commonly compromised services}

The top 10 services that are most frequently compromised in \INFECTIONS{} are:
google\-.com,
facebook\-.com,
live\-.com,
roblox\-.com,
twitter\-.com,
discord\-.com,
twitch\-.tv,
epicgames\-.com,
instagram\-.com,
and paypal\-.com.

Because each of these is a well-known web service (c.f.\ Similarweb's ``Most Visited Websites Ranking Analysis''\footurl{https://www.similarweb.com/top-websites/}),
the results are consistent with expectations.

\begin{figure}[tb]
\centering
\includegraphics[width=\linewidth]{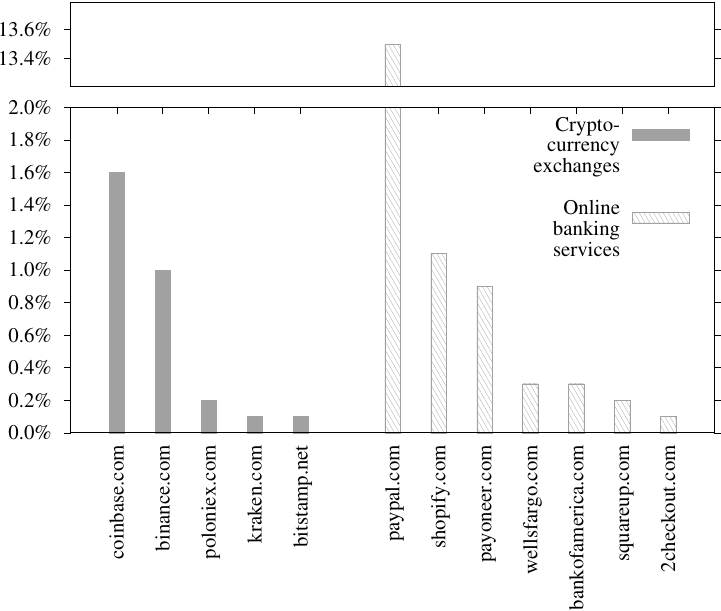}
\Description[Financial services]{Financial services, cryptocurrency exchanges and online banking.}
\caption{Financial services in \INFECTIONS{}:
cryptocurrency exchanges and online banking.
13.5\% of the victims have their PayPal online payment account compromised.}%
\label{fig:financial}
\end{figure}

\subsection{Compromised financial services}

The victims used the financial services listed in \autoref{fig:financial},
consisting of online banking services and cryptocurrency exchanges.
In total, 13.5\% of the victims in \INFECTIONS{} have their PayPal online payment account compromised.

\subsection{Victim records for sale}

Adversaries may attempt to profit from stolen credentials, e.g.,
by the owner of the bot network using them,
or the adversary selling the access information forward.
We now provide evidence of the latter with our matching criteria described in \autoref{sec:methods} (methods).

We first automatically extract and identify unique domain names (compromised services) from \VICTIMS{},
and select records that contain at least five of these distinct domain names.
This selection yields 2,116 records from \VICTIMS{}.
We then search for matches from \INFECTIONS{}.

As an example of our selection criteria, consider 1502766.json from \INFECTIONS{} and 1814801285.json from \VICTIMS{}.
Here, countries (IT and IT) match, and both records contain a total of 11 distinct services:
gravel\-.ltd,
bitminer\-.biz,
shamining\-.com,
coincity\-.in,
navminer\-.biz,
luckfarm\-.fun,
ethereum-master\-.com,
aristabank\-.com,
flovin\-.ltd,
getdoge\-.io,
and xcryptos\-.io.

As a result, we discover that 5\% (106/2,116) of the malware-infected victims are present in both datasets.
Please note that this method only provides evidence that infostealer malware victims are available for purchase on Genesis Market; the exact percentage must be higher than this.
This unequivocally establishes a link in the supply chain:
Infostealer malware infections are a supplier for cybercriminal marketplaces that sell access to victim data.

\subsection{Access to a malware victim: ask prices}

\textit{\autoref*{rq:device}: How are the prices determined for compromised devices?}

Genesis Market sells access to victim information, with stolen credentials from victims serving as the primary pricing factor.
Addressing \autoref{rq:device}, we discover that the number of compromised services seems to dictate the ask price.
Using linear regression to determine the price-to-services correlation, we obtain:
\begin{displaymath} Price = 0.05*Services + 6.26 \end{displaymath}

Interestingly, there is no effect of country on pricing.
We determined this by creating separate linear regression models for each nation, then comparing them.

\begin{figure}[tb]
\centering
\includegraphics[width=\linewidth]{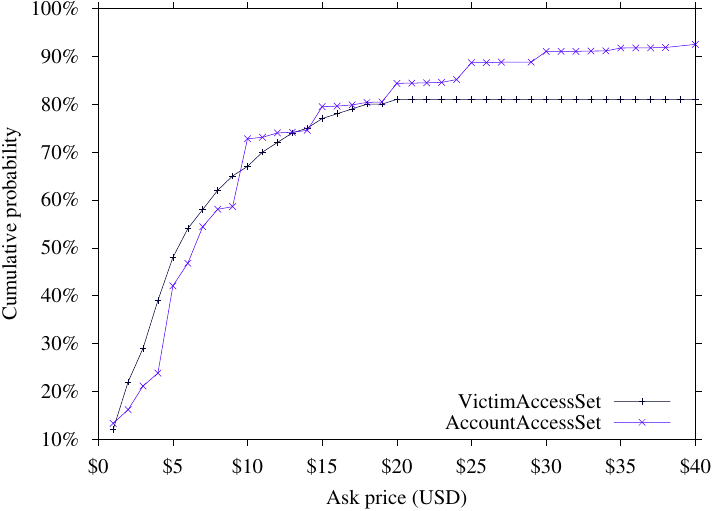}
\Description[Price of access]{Price of access to compromised devices and accounts.}
\caption{Price of access to compromised devices (\VICTIMS{}) and accounts (\COMPROMISE{}).
In the former, 91\% fall between 1 and 20 US dollars, with a median of 5 US dollars.
In the latter, 91\% fall between 1 and 30 US dollars, with a median of 7 US dollars.}%
\label{fig:prices}
\end{figure}

Finally, we examine price distribution to conclude our study of \autoref{rq:device}.
Rounding the prices to the nearest dollar,
the range of prices is from 1 to 350 US dollars,
and prices over 125 US dollars each have fewer than 10 victims.
\autoref{fig:prices} illustrates these ask prices, where 91\% fall between 1 and 20 US dollars, with a median of 5 US dollars.

\subsection{Access to an online account: ask prices}

\textit{\autoref*{rq:accounts}: Compared to access to a victim's device, what is the cost of online account access?}

\COMPROMISE{} reveals the prices of compromised online account access,
where we chose a subset of 599 account advertisement pages.
We manually confirmed these contain individual accounts with price information, and no other items for sale,
such as accounts combined into a single item, organisation databases, or stolen software.

\autoref{fig:prices} illustrates these ask prices, where 91\% fall between 1 and 30 US dollars, with a median of 7 US dollars.
This data provides quantitative estimates for the costs of individual online accounts, addressing \autoref{rq:accounts}.

In summary,
after purchasing access to a victim's device for 1--20 US dollars,
a cybercriminal is able to verify working online accounts of the victim
and resell the harvested online accounts for 1--30 US dollars per account.

\subsection{Malware supply chain}

\begin{figure}[tb]
\centering
\includegraphics[width=\linewidth]{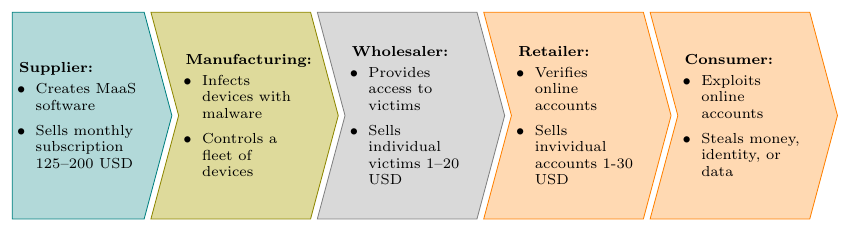}
\Description[Malware value chain]{Malware value chain: malware suppliers, infections, victim sales, and online account sales.}
\caption{Malware value chain: malware suppliers, infections, victim sales, and online account sales.}%
\label{fig:valuechain}
\end{figure}

Malware suppliers, infections, victim sales, and online account sales form the value chain in \autoref{fig:valuechain}.

Malware victim and account access are both perishable goods, so the manufacturer must sell them before they grow stale.
To sell victims, the manufacturer needs
economies of scale (roughly, an economics concept formalising cost advantages)
and a streamlined production process.
In \autoref{tab:profit}, we display the known costs and projected earnings for the operations.

\begin{table*}[htb]
\Description[Malware value chain]{Prices and profitability in the malware economy value chain.}
\caption{Prices and profitability in the malware economy value chain.
Many goods, such as licences, have fixed costs and zero per-unit production costs at scale.
The average number of compromised services for victims in the Genesis Market is 48.}\label{tab:profit}
\resizebox{1.0\linewidth}{!}{%
\begin{tabular}{p{0.2\linewidth}p{0.21\linewidth}p{0.17\linewidth}p{0.42\linewidth}}
  \textbf{Actor} & \textbf{Price (US dollars)} & \textbf{Pricing model} & \textbf{Monthly profit is revenues minus total costs} \ \\
  {Malware-as-a-Service} & 125--200 / monthly license & Monthly subscription &
  \begin{math} \sum {Pay\textrm{-}per\textrm{-}subscriptions} \end{math} \\
  {Spam-as-a-Service} & 100--500 / million sent emails & {Pay-per-million} &
  \begin{math} \sum {Pay\textrm{-}per\textrm{-}million\textrm{-}emails} \end{math} \\
  {Email Target List Supplier} & 25--50 / million addresses & {Pay-per-million} &
  \begin{math} \sum {Pay\textrm{-}per\textrm{-}million\textrm{-}addresses} \end{math} \\
  {Infection Manufacturing} & Bulk sale of victims & {Pay-per-victim-access} &
  \begin{math} \sum ({Pay\textrm{-}per\textrm{-}victim\textrm{-}access}-comission) \end{math} \\
  {Access Wholesaler} & 1--20 / victim access & Commission pricing (?) &
  \begin{math} \sum {Pay\textrm{-}per\textrm{-}commission} \end{math} \\
  {Account Retailer} & 1--30 / stolen record & {Pay-per-access} &
  \begin{math} \sum ({Pay\textrm{-}per\textrm{-}account\textrm{-}access}-Pay\textrm{-}per\textrm{-}victim\textrm{-}access/48) \end{math} \\
  {Criminal Consumer} & Exploits individual accounts & {Per-victim-revenue} &
  \begin{math} \sum ({Account\textrm{-}revenue}-{Pay\textrm{-}per\textrm{-}account\textrm{-}access}) \end{math} \\
\end{tabular}
 }
\end{table*}

\subsubsection*{Cause interruption.}
Victim wholesale is precisely what Genesis Market offers.
Wholesale is the single weakest link in the chain:
As access to victim devices is a single point of failure through one marketplace,
law enforcement should focus effort there.
\section{Limitations and quality metrics}\label{sec:quality}

\subsection{\INFECTIONS{}: representativeness}

\begin{figure}
\centering
\includegraphics[width=\linewidth]{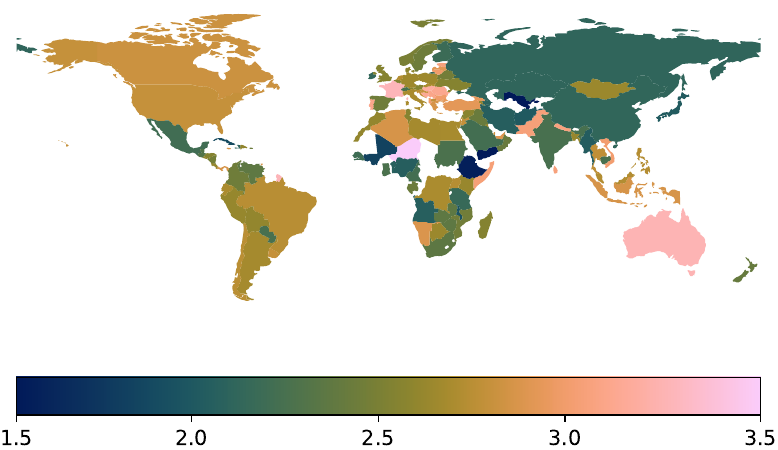}
\Description[Infections in countries]{The number of infections per million Internet users, using a base-10 logarithmic scale.}
\caption{Countries in \INFECTIONS{} by the number of victims per million Internet users, using a base-10 logarithmic scale.}%
\label{fig:countries}

\phantom{ZZ}

\includegraphics[width=\linewidth]{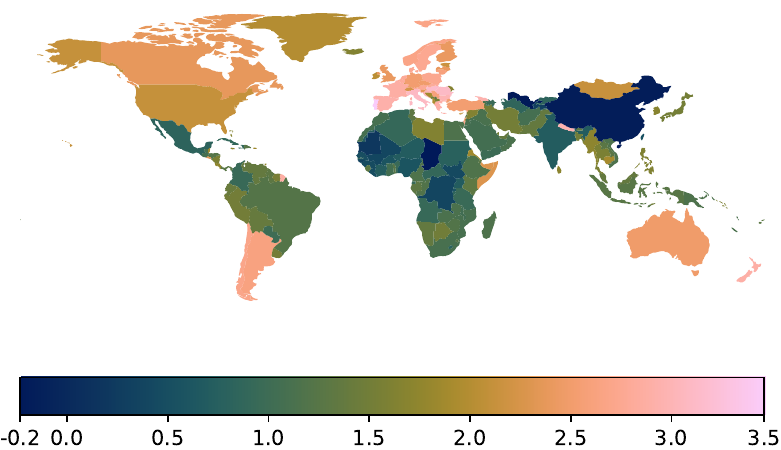}
\Description[Victims in countries]{The number of Genesis Market victims per million Internet users, using a base-10 logarithmic scale.}
\caption{Genesis Market sells access to victims.
Countries in \VICTIMS{} by the number of victims per million Internet users, using a base-10 logarithmic scale.}%
\label{fig:countries_genesis}
\end{figure}

Here we study the geographic distribution of entries in \INFECTIONS{}.
The World Bank publishes open access to world statistics\footurl{https://doi.org/10.1596/978-1-4648-0382-6_open_data},
including estimated number of Internet users.
This statistic allows us to calculate the proportion of victims per million Internet users
(International Telecommunication Union (ITU) World Telecommunication/ICT Indicators Database, CSV file)\footurl{https://doi.org/10.5281/zenodo.3969472}.
We compared the number of victims to that estimate for each country.
\autoref{fig:countries} illustrates the country distribution across this dataset,
which ranks countries by the number of victims per million Internet users.

Although the distribution of victims varies between countries, nearly every country has victims.
The number of victims may vary due to the country- or language-specific nature of some infection sources.
For example, Niger, France, and Australia have a disproportionately high number of cases in comparison to their Internet user populations.

\INFECTIONS{} contains compromised credentials for 424 unique onion domains from 293 victims,
i.e.\ addresses from TorBrowser for use within the Tor anonymity network.
We opened onion domains to view them: Most of them are offline,
but the online onion services appear to be marketplaces, based on the landing page.
Typically, marketplaces on Tor sell illegal substances and goods \cite{un2020}.

The Tor Network has an estimated 3 million concurrent users, based on daily data collection of the Tor Project for 2022\footurl{https://metrics.torproject.org/userstats-relay-country.html}.
This equates to 0.075\% of the total 4 billion Internet users\footurl{https://data.worldbank.org/indicator/NY.GDP.PCAP.CD?most_recent_value_desc=true}.
We now compare this percentage to what our dataset indicates.
Using \INFECTIONS{}, we are aware of the percentage of victims who authenticate on onion websites (the total number of victims using Tor is greater than this), which is 0.016\% (293/1.8 million).
The proportions have the same order of magnitude, and \INFECTIONS{} also represents anticipated Tor users.

According to a Google study in conjunction with Harris Poll in 2018,
13\% of US adults use a single password for all of their accounts\footurl{https://services.google.com/fh/files/blogs/google_security_infographic.pdf}.
We conducted a similar study using \INFECTIONS{},
and \autoref{fig:passwords_usernames_emails} depicts the results.
We observe that 13\% of victims use a single password for all services,
strikingly consistent with the aforementioned study.

In summary, it is fair to say \INFECTIONS{} represents global Internet users.

\subsection{\INFECTIONS{}: accuracy}

We compare \INFECTIONS{} to other data sources, including the number of active Facebook users.
Given that there are 2.9 billion monthly active users, Facebook estimates that 60\% of the world's active Internet users
log in each month in 2021\footurl{https://investor.fb.com/investor-news/press-release-details/2022/Meta-Reports-Fourth-Quarter-and-Full-Year-2021-Results/default.aspx}.
In \INFECTIONS{}, 46\% (829,630/1,809,384) of victims had their Facebook accounts compromised (matching in services:*.facebook.com).

\subsubsection*{Multi-victim entries.}
It is possible that the infected device is a shared computer with dozens of users in a library, school, or business.
In \INFECTIONS{},
3.3\% of entries (60,163)
have over ten distinct email addresses, over 50 passwords, and over 100 compromised services.
As such, multiple victims are almost certainly represented by these particular entries.
However,
in 96.7\% of cases it appears as though one infected device equates to one human victim in \INFECTIONS{}.

\subsubsection*{Not all timestamps are accurate.}
The primary reason is that text extractors are interpreting the victim's information incorrectly.
For instance, we noticed that a victim has a computer named 2011-01-1, and because this information is near the actual time field, the extractor chose it.
The second reason is that the timestamp appears to be incorrect (perhaps it uses the victim's clock time).

\subsubsection*{Empty victim entries.}
0.6\% of entries (11,157) contain no detailed victim information.
This indicates that the extractors did not discover any timestamps, IP addresses, domain names, or email addresses.
We retain these empty entries in \INFECTIONS{} because they represent victims that had no meaningful data stolen.

\subsubsection*{Anti-virus information is only indicative.}
Although the anti-virus field contains snippets of potential endpoint protection software,
it does not always show if the device is running the most recent protection.
This field is also populated if the victim is merely a customer of a security company
(e.g., credentials for my\-.kaspersky\-.com).
Consequently, we do not attempt to quantify the effectiveness of endpoint protection with our dataset.

\subsubsection*{Malware also stores incorrect credentials for services.}
In the raw data, there are numerous credentials for the same service
where the username (which can be an email address) and password vary slightly.
When victims enter their credentials incorrectly, malware obliviously stores these incorrect credentials for services.
For instance, it is common for victims to have usernames and email addresses like name@gmail.com and name@gmail.co.

\subsection{\VICTIMS{}: representativeness}

Here we study the geographic distribution of entries in \VICTIMS{}.
\autoref{fig:countries_genesis} illustrates the country distribution across this dataset,
which ranks countries by the number of victims per million Internet users
(i.e., analogous to \autoref{fig:countries}, but for \VICTIMS{}).

According to \autoref{fig:countries_genesis}, nearly every country is in \VICTIMS{}.
Although the distribution of victims varies between countries, nearly every country has victims.
The number of victims may vary due to the country- or language-specific nature of some infection sources.

For example, Portugal, Hungary, Romania, Italy, and Bulgaria have a disproportionately high number of cases in comparison to their Internet user populations.
In contrast, Uzbekistan (post-Soviet state), China, and Tajikistan (post-Soviet state) have a disproportionately low number of cases.

Certain countries do not appear in \VICTIMS{} at all:
For example, the lack of Russian victims is clear from \autoref{fig:countries_genesis}.
This indicates that Genesis Market operates from Russia,
where cybercriminals are actually allowed to operate freely as long as they do not target Russian citizens.
Furthermore, this explains the two post-Soviet state outliers previously mentioned,
as well as the following post-Soviet states that are absent from \VICTIMS{}:
Ukraine, Belarus, and Kazakhstan.
Lastly, North Korea is also absent from \VICTIMS{}.

This marketplace sells access to devices in all major countries except Russia (and several post-Soviet states),
and indeed the Genesis Market domain resolved to an IP address there (before the FBI seized it in 2023).

Authorities in Russia will not investigate cybercrime
unless a Russian company or person files a formal complaint as a victim \cite{schneier2021}.
The simplest way for these criminals to remain undetected is to ensure
that their malware does not infect people in their home country.
For instance, DarkSide malware does not spread within the Commonwealth of Independent States (CIS) or
former Soviet satellites,
and many malware strains do not infect computers with a Cyrillic keyboard \cite{schneier2021}.

In summary, it is fair to say \VICTIMS{} represents global Internet users, sans the noted omissions.

\subsection{\VICTIMS{}: accuracy}

Here we describe a number of validations to demonstrate that \VICTIMS{}
accurately represents global Internet users
(by comparison to anticipated numbers).

\subsubsection*{Compromised services.}
In \VICTIMS{}, 50\% (239,622/480,994) of victims had their Facebook accounts compromised.
This corresponds to the expected number of Facebook users in the dataset,
as Facebook estimates that 60\% of the world's active Internet users log in each month.

\subsubsection*{Operating systems.}
Windows 10 has a 72.23\% market share among Windows operating systems,
according to Statcounter GlobalStats in 2022\footurl{https://gs.statcounter.com/os-version-market-share/windows/desktop/worldwide}.
We calculate the percentage of Windows 10 users within our \VICTIMS{}.
According to our data, the proportion of Windows 10 users is directly analogous, at 76\% (365,676/480,994).

\subsubsection*{Web browsers.}
The Chrome web browser has a market share of 79.9\% according to W3Schools' Browser Statistics for 2022\footurl{https://www.w3schools.com/browsers/default.asp}.
Likewise, 71\% (340,933/480,994) of the victims in \VICTIMS{} are Chrome users.
\section{Research ethics and privacy}\label{sec:privacy}

Open Data is an unqualified good, but sometimes it can identify individuals.
For example,
\citet{tran2015open} suggest that avoiding the collection of certain types of data can protect against civil rights violations,
and countries with a history of data misuse have resolved not to collect data that associates an individual with a vulnerable group.

We obtained the original information for our research from public sources,
even though some sources are difficult to locate and access.
Bear in mind that at least some of the original logs have been available online for three years.
Worse yet, as we demonstrated in \autoref{sec:results},
this information is partially available for sale.
As a result, an adversary can locate the original information and identify the victims from the original data,
as their real names are typically included in the available data logs.

Our research results use anonymised victim data, which contains no personal information.
The original data sources for \VICTIMS{} and \COMPROMISE{} are anonymous and do not contain personal information,
hence require no protection.
The original data logs for \INFECTIONS{} contain private and sensitive information.
We process \INFECTIONS{} while protecting the anonymity of victims (see \autoref{sec:methods}).

Malevolent actors would simply use the original data source, not our datasets.
By making our structured datasets publicly available with privacy protections in place,
researchers can study malware without accidental exposure to sensitive victim information,
nor the security risks of collecting the original data.
\section{Conclusion}\label{sec:conclusion}

We now summarise the entire infostealer malware ecosystem of actors and their operations.
We find that cybercrime is specialising in several roles within the illicit economy and for the first time give a comprehensive illustration of the malware value chain with exact pricing information.

First, unethical software developers create malware. They sell software licenses via the Internet,
where the cost of the MaaS model is between 125--200 US dollars per month.

Second, the owner of the malware network infects victims via popular methods,
such as email attachments and macros.
This attacker may utilise spam-as-a-service,
which costs between 100--500 US dollars per million emails sent, and targeted email lists,
which costs between 25--50 US dollars per one million email addresses.

Third, once malware infects victims,
the network operator sells access to their devices on Genesis Market.

One key new finding is a full understanding of the precise monetary value of victims from a criminal perspective:
The price range is between 1--350 US dollars,
and the value of a victim's record depends on the number of compromised services.
This provides insight into the malware network owner's direct retail value of the victim's information.
Typically, access to a malware victim costs between 1--20 US dollars, with a median of 5 US dollars.

Fourth and finally,
a buyer of victim access ultimately profits by gaining access to the victim's online accounts or by reselling individual credentials.
Database Market operates inside the anonymous Tor network,
and resells access to compromised online accounts.
We show how resale prices are typically between 1--30 US dollars, with a median of 7 US dollars.
\section{Future work}\label{sec:future}

\subsubsection*{Use datasets for interdisciplinary research.}
With our datasets, it is possible to investigate completely new research topics outside of security and malware.
We provide one example here: studying the effectiveness of Internet censorship.
In several nations, including the Philippines, the popular adult website
Pornhub is blocked\footurl{https://www.cnnphilippines.com/news/2017/01/16/government-blocks-major-porn-websites.html}.
Although using the service without creating an account is free, additional paid features are available after logging in.
We investigate this using \INFECTIONS{}, and provide a Python script for this with the published data.
0.47\% (133/27,978) of victims in the Philippines have a premium account on Pornhub although the service is censored.
Comparatively, 1\% (1,861/1,864,61) of victims in the United States and 0.2\% (108/56902) of victims in Turkey have premium Pornhub accounts.
As a way to use these datasets further, we just showed that Internet filtering in the Philippines does not prevent many people from using a popular adult website.

\subsubsection*{User guidelines and security technologies.}
We propose a study of novel guidelines and techniques for businesses that should account for the fact
that infostealer malware undermines multi-factor authentication (MFA) and endpoint security,
thereby rendering the end-user defenceless.
If we compare our malware victim data to the National Institute of Standards and Technology's (NIST) guidelines for the prevention and management of malware incidents,
we find that infostealer malware undermines these suggestions \cite{mell2005guide,souppaya2013guide}.
According to the guidelines,
network-based anti-virus software, host-based anti-virus software, and spyware detection and removal software
are the most effective approaches for preventing malicious code \cite[p.\ 77]{mell2005guide}.
A typical victim uses Windows 10, which by default has Microsoft's most recent endpoint defences.
Microsoft Office macros remains a common infection method, after decades of security patches against these exploits.
A typical victim has 1--8 unique passwords,
indicating that the typical user finds it too challenging to use a unique password for each account.
A user may employ MFA,
but Genesis Market duplicates the victim's authenticated web session,
hence weakening MFA protection.

\begin{acks}
This project has received funding from the European Union's Horizon 2020 research and innovation
programme under grant agreement No 952622 (SPIRS),
and grant agreement No 804476 (SCARE).
\end{acks}
 
\bibliographystyle{ACM-Reference-Format}
\interlinepenalty=10000

\end{document}